# Enhanced Gas-Flow-Induced Voltage in Graphene

Jun Yin, Jianxin Zhou, Xuemei Li, Yaqing Chen, Guoan Tai and Wanlin Guo[*]

Key Laboratory for Intelligent Nano Materials and Devices (MOE) and State Key Laboratory of Mechanics and Control of Mechanical Structures, Nanjing University of Aeronautics and Astronautics, Nanjing 210016, China

**ABSTRACT:**

We show by systemically experimental investigation that gas-flow-induced voltage in monolayer graphene is more than twenty times of that in bulk graphite. Examination over samples with sheet resistances ranging from 307 to 1600 Ω/sq shows that the induced voltage increase with the resistance and can be further improved by controlling the quality and doping level of graphene. The induced voltage is nearly independent of the substrate materials and can be well explained by the interplay of Bernoulli's principle and the carrier density dependent Seebeck coefficient. The results demonstrate that graphene has great potential for flow sensors and energy conversion devices.

[*]e-mail: wlguo@nuaa.edu.cn

Use of wind power and creation of self-powered devices are most attractive issues in both academic and industry fields.[1-3] Gas flow passing the surface of a conductive material can generate a voltage due to the interplay of Bernoulli's principle and the Seebeck effect.[4] Graphene, as the thinnest material with extraordinary electrical properties and flexibility, strong mechanical strength,[5-9] can be split off from bulk graphite by mechanical or chemical cleavage methods,[10,11] directly grown on large scale on surfaces of many materials.[12-15] It can be transferred to various substrates, transforming the surfaces from insulator into conductor[14-17] and protecting the surfaces from oxidation and corrosion.[18] If all these promising properties can be merged with gas-flow-induced voltage, graphene will have great potential for velocity mensurement, wind energy conversion and creation of self-powered devises. However, the bulk graphite shows poor ability to yield voltage from gas flow.[4] Here, we demonstrate that the single layer graphene can enhance the ability in produce voltage from gas flow by twenty-fold over the multilayered graphite, nearly independent of the substrates. This intriguing ability can be further improved by controlling the quality and carrier concentration, thus opens new potential for flow sensors, self-powered devices and energy conversion.

Single layer graphene samples used in this work was synthesized by recently developed chemical vapor deposition (CVD) method.[13-15] After growth, the CVD graphene on the copper foil was firstly transferred on to a 1 mm thick quartz substrate.[17] A photo of a graphene sheet of 5×25 mm$^2$ in size transferred onto the quartz substrate is shown in the inset of Fig. 1(a). The transferred graphene samples were then characterized by a Renishaw Raman spectrometer with solid-state laser ($\lambda$ = 514.5 nm), as shown by Fig. 1(a). The well symmetric 2D peak shape with full width at half-maximum (FWHM) ~31 cm$^{-1}$, small ratio of G to 2D peak intensity ($I$(G)/$I$(2D) < 0.5) and negligible disorder induced D peak clearly show that the sample is monolayer graphene with low density of defects.[19,20] The blueshifted position of G peak (~1590 cm$^{-1}$) for our sample compared to that of exfoliated graphene (~1586 cm$^{-1}$) indicates relatively higher carrier concentration (>3×10$^{12}$/cm$^2$).[21] The gas-flow-induced voltage ($V_{GF}$) of the sample was measured using devices consisting of uncovered graphene sheet, copper electrodes and directed flow pipe as schematically illustrated in Fig. 1(b). Both the electrical contacts were made with copper foil adhered by silver emulsion. The exposed part of the graphene sample was adjusted to 10 mm long along the flow direction and 5 mm wide perpendicular to the flow. The outlet of gas flow tube is kept 5 mm away at an angle $\alpha$ = $\pi/4$ with respect to the graphene surface to obtain a maximum output voltage signal. The sheet resistance of the sample is measured to be 943 $\Omega$/sq. The velocity of the gas flow ($v$) was measured using a rotameter and a digital pressure indicator. The induced voltage was measured using KEITHLEY 2010 multimeter.

Figure 1(c) shows the typical voltage response from the above setup system as a function of time when the gas flow is switched on and off. The response of a graphite sample (highly oriented pyrolytic graphite, HOPG) is also displayed for comparison. When an Ar flow with velocity of 68 m/s is turned on, the induced voltage signal increases from zero to -58 μV and 2.5 μV for graphene and graphite, respectively.

Apparently, the signs of $V_{GF}$ for graphene and graphite are opposite, and the magnitude of $V_{GF}$ for graphene is over twenty-fold larger than that for graphite of comparable size. As the gas flow is turned off, the voltage decreases, and eventually returns to zero. The alternating on-off gas flows produce a series square-wave-like voltage signals in the sample and the voltage response to the gas flow is within a few seconds.

To examine the influence of sample quality of graphene for the flow induced voltage, we performed the measurement with 12 CVD graphene samples transferred onto quartz substrate, with sheet resistance values ranging from 425 $\Omega$/sq to 1.6 k$\Omega$/sq. Figure 2(a) shows the variation of $V_{GF}$ for the graphene devices against $\sigma$, the reciprocal of sheet resistance. It can be seen that the $V_{GF}$ presents slightly increasing tendency with the value of sheet resistance. According to the theory developed in carbon nanotube and other materials,[4] the $V_{GF}$ can be deduced as $V = kS\Delta T$, where $S$ is the Seebeck coefficients of the sample material with respect to the electrode material copper, $\Delta T$ is the gas flow induced temperature difference along the sample, $k$ is a fitting factor. In the case of graphene devices, the divergent values of $V_{GF}$ recorded at the same gas-flow condition suggest the different Seebeck coefficients for various graphene samples. The negative sign of $V_{GF}$ suggests a *p*-type doping state, in consistent with literature reports that graphene can be easily *p*-doped via absorption of water molecular or oxidative ions onto its surface in ambient air.[22-24]

The Seebeck coefficients of graphene is strongly dependent on the carrier density *n*,

theoretical calculations based on experimental results.[25-27] and the effective-medium theory[28] show that the absolute value of $S$ for graphene peaks at $n \sim 1\times 10^{12}/cm^2$ and decreases with further increase in $n$. In high density region, the Seebeck coefficient of graphene behaves as $1/\sqrt{n}$, in agreement with the semiclassical Mott formula. As the transferred CVD graphene typically has carrier density larger than $1\times 10^{12}/cm^2$ due to the doping of charged impurities in transfer process,[15] which is also confirmed by our Raman result, and the lower sheet resistance of graphene film implies higher carrier density, the decreasing tendency of $V_{GF}$ against increasing $\sigma$ in Fig. 2(a) can be well explained by the density dependence of Seebeck coeffiecient in graphene.

The carrier density of graphene can be effectively controlled by doping level in graphene. To modulate the concentration of $p$-type carriers (holes), the graphene sample was treated in a nitric acid ($HNO_3$) vapor environment[15] for 15 s, 60 s and 180 s, the correspongding I-V curves for the pristine and treated sample are shown in the inset of Fig. 2(b). The sheet resistance reduced from 943 Ω/sq for pristine graphene, to 307 Ω/sq for the sample with 180 s treatment. Following each $HNO_3$ treating process, $V_{GF}$ measurements were conducted under an argon gas flow at velocity of 68 m/s. The square-wave-like voltage responses are shown in the inset of Fig. 2(b). The deeper $p$-doped graphene sample with lower resistance has a weaker response than the pristine sample, and the induced voltage decreases from 58.5 μV to 38.5 μV gradually. The $V_{GF}$ against $\sigma$ from our experiments, and calculated $S$ against $n$ by effective-medium theory (near charge-neutrality point) and Mott formula (high carrier density region)[27] are plotted in Fig. 2(b), where the experimental data shows very similar tendency to the theoretical curve. Note that although $V_{GF}$ is demonstrated to be reduced by $HNO_3$ vapor treatment here, it's realizable that $V_{GF}$ can be enlarged by reducing the carrier density to a proper value.

The gas-flow-induced voltage $V_{GF}$ of various materials is strongly flow-velocity-dependent. The dependence of voltage signal on the square of Mach number $M$ for graphene and graphite is shown in Fig. 3(a), where $M = v/c$, $v$ is gas flow velocity, $c$ is the sound velocity in medium (323 m/s for argon at 300 K). It can be seen that the value of $V_{GF}$ increases linearly with $M^2$, with slop of -1325 for graphene and 56 for graphite. The slope $A$ has been found to meet the relationship $V/M^2 \propto A\gamma S$, where $\gamma$ is the heat capacity of gas.[4] Since the same gas type used in the measurements, the larger magnitude of $V_{GF}$ for graphene comes from the larger Seebeck coefficients, which is supported by recently reports about the Seebeck coefficient of graphene.[22-25] The flow-velocity dependence of $V_{GF}$ for graphene in argon, nitrogen and oxygen gases exhibit similar linear tendency (Fig. 3(b)), except the slightly different slopes which is mainly attributed to the difference in $\gamma$ value [$\gamma$ (nitrogen) : $\gamma$ (oxygen) : $\gamma$ (argon) =1 : 1 : 1.2 ]. The lower slops at high velocity region ($M^2 > 0.05$) are attributed to the density changes of the gas.

As a flexible, stable and easy-processable material, graphene can be transferred to varied substrates including polymers and crystal wafers. To find the influence of substrates, we measured the gas flow induced $V_{GF}$ of graphene samples of the same size on different substrates under argon gas flow of 68 m/s. The average output signal is measured to be 53 µV for polytetrafluoroethylene (PTFE), 50 µV for polymethylmethacrylate (PMMA), 53 μV for poly(ethylene terephthalate) (PET), 55 μV for quartz, 51 μV for mica and 54 μV for 300 nm $SiO_2$/Si substrates. As the measurement error is about ±4 μV, the measured range for different substrates suggests that the $V_{GF}$ response of graphene is nearly independent of the insulating substrates.

In conclusion, we have shown that the gas flow induced voltage in graphene can be more than twenty times of that in bulk graphite, owing to its atomic-thin structure with high Seebeck coefficient. The induced voltage presents increasing tendency with the sheet resistance, which can be explained by the carrier density dependent Seebeck coefficient. The high ability of graphene to produce voltage under gas flow is proven to be robust to substrate materials and can be further improved by controlling the quality and doping level. The results exhibit that graphene can be a promising candidate for gas flow sensors and high efficient energy conversion devices.

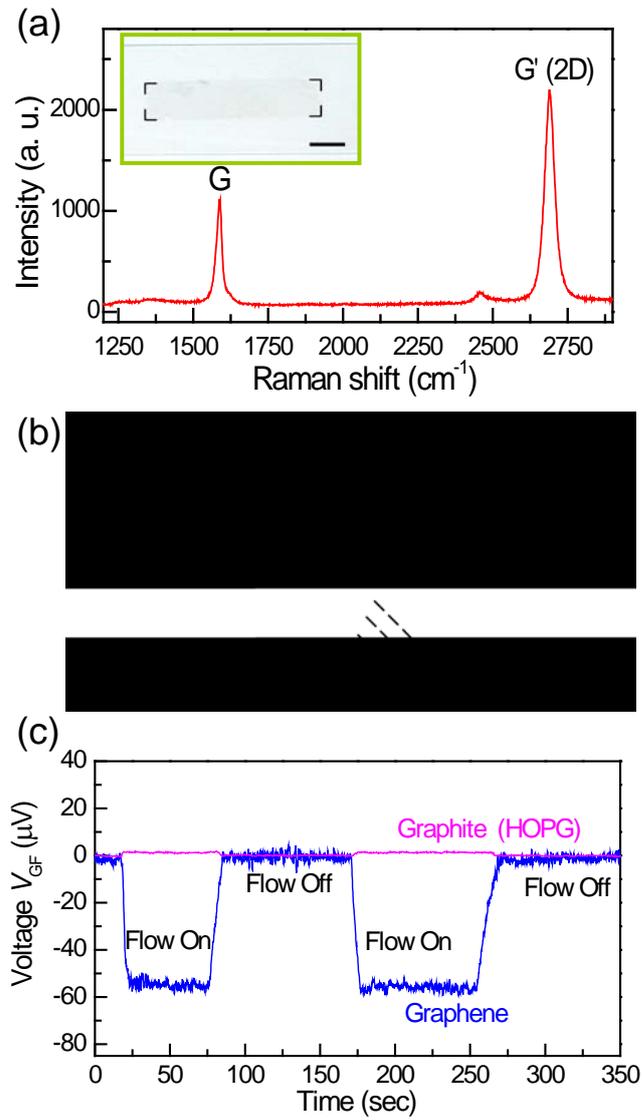

FIG. 1. (Color online) (a) Typical Raman spectra of single layer graphene on quartz substrate, the inset shows a photo of graphene sample, bar = 5 mm. (b) The schematic illustration of the gas flow experiment setup. (c) Typical voltage signal response of graphene and graphite for switching on/off the argon gas flow at 68 m/s.

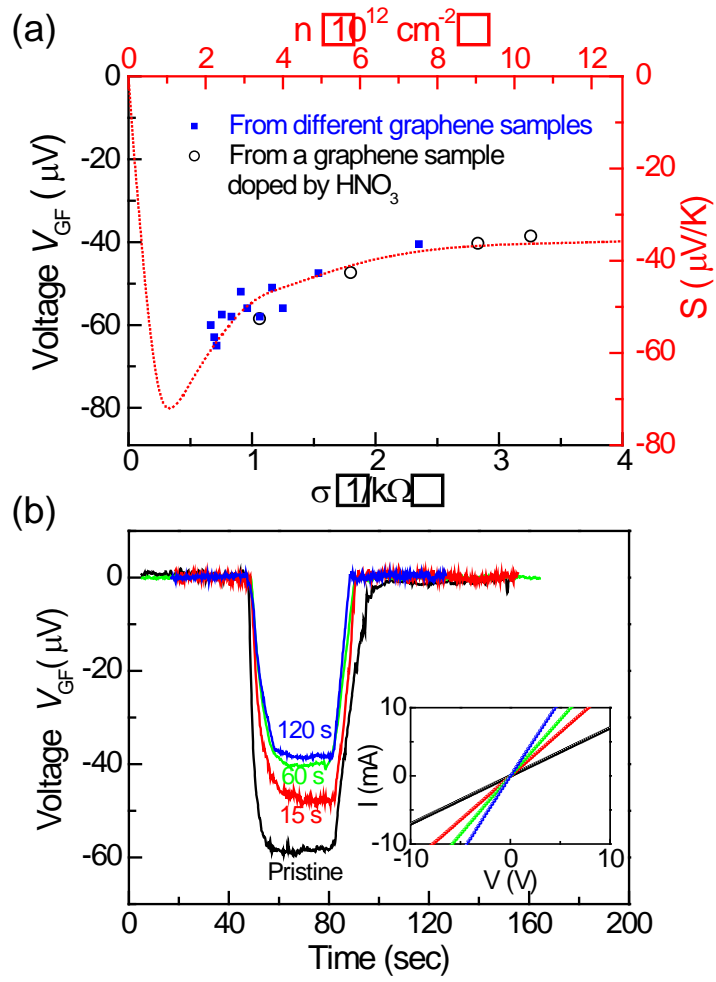

FIG. 2. (Color online) (a) Voltage response for graphene samples with different resistances (blue squares) and a graphene sample with different HNO$_3$ treating durations (black circles). Doted line shows the predicted Seebeck coefficient $S$ as a function of carrier concentration at 300 K following the effective-medium theory.[28] (b) Voltage response of pristine and the HNO$_3$ treated graphene samples in argon gas flow at 68 m/s. Insert shows $I$-$V$ curves of the samples.

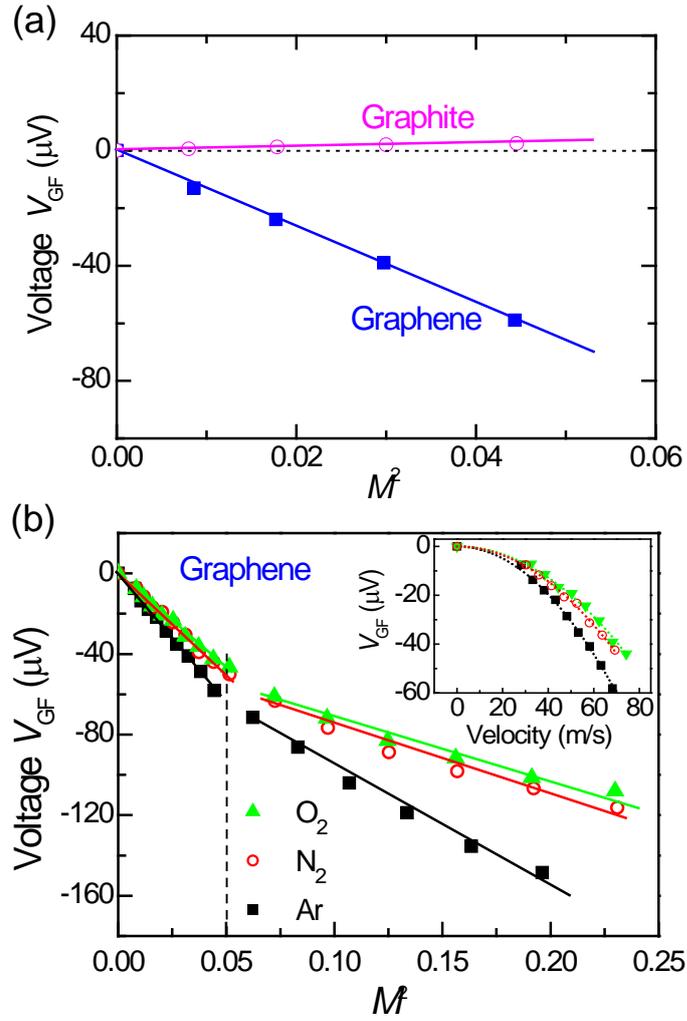

FIG. 3. (Color online) (a) Ar gas flow induced voltage $V_{GF}$ versus $M^2$ for graphene and graphite. (b) Variation of $V_{GF}$ against $M^2$ for argon, nitrogen and oxygen gases. Solid lines are linear fitted to the data. Insert of (b) shows the dependence of $V_{GF}$ on velocity at low velocity region.